\documentclass[aps,pra,twocolumn,amsmath,amssymb,nofootinbib,showpacs,superscriptaddress]{revtex4-1}
\usepackage[english]{babel}
\usepackage{latexsym}
\usepackage{graphics}
\usepackage{graphicx}
\usepackage{epsfig}
\usepackage{color}
\usepackage{bm}
\usepackage{amsmath}
\usepackage{amssymb}
\usepackage{amsthm}
\usepackage{dcolumn}
\usepackage{bm}
\usepackage{float}
\usepackage{hyperref}
\usepackage{color}
\usepackage{epstopdf}
\usepackage{cleveref}
\usepackage[svgnames]{xcolor}
\usepackage{soul}
\usepackage{multirow}
\hypersetup{hidelinks,colorlinks=true,allcolors=DarkBlue}
\usepackage{ulem}

\usepackage{soul}

\theoremstyle{remark}

\begin{document}
\setstcolor{red}

\preprint{APS/123-QED}

\title{Genome assembly using quantum and quantum-inspired annealing}

\author{A.S. Boev}
\affiliation{Russian Quantum Center, Skolkovo, Moscow 143025, Russia}

\author{A.S. Rakitko}
\affiliation{Genotek ltd., Moscow 105120, Russia}

\author{S.R. Usmanov}
\affiliation{Russian Quantum Center, Skolkovo, Moscow 143025, Russia}

\author{A.N. Kobzeva}
\affiliation{Russian Quantum Center, Skolkovo, Moscow 143025, Russia}

\author{I.V. Popov}
\affiliation{Genotek ltd., Moscow 105120, Russia}

\author{V.V. Ilinsky}
\affiliation{Genotek ltd., Moscow 105120, Russia}

\author{E.O. Kiktenko}
\affiliation{Russian Quantum Center, Skolkovo, Moscow 143025, Russia}
\affiliation{Moscow Institute of Physics and Technology, Dolgoprudny 141700, Russia}

\author{A.K. Fedorov}
\affiliation{Russian Quantum Center, Skolkovo, Moscow 143025, Russia}
\affiliation{Moscow Institute of Physics and Technology, Dolgoprudny 141700, Russia}

\date{\today}
\begin{abstract}
Recent advances in DNA sequencing open prospects to make whole-genome analysis rapid and reliable, which is promising for various applications including personalized medicine.
However, existing techniques for {\it de novo} genome assembly, which is used for the analysis of genomic rearrangements, chromosome phasing, and reconstructing genomes without a reference, require solving tasks of high computational complexity. 
Here we demonstrate a method for solving genome assembly tasks with the use of quantum and quantum-inspired optimization techniques.
Within this method, we present experimental results on genome assembly using quantum annealers both for simulated data and the $\phi$X 174 bacteriophage.
Our results pave a way for an increase in the efficiency of solving bioinformatics problems with the use of quantum computing and, in particular, quantum annealing.
We expect that the new generation of quantum annealing devices would outperform existing techniques for {\it de novo} genome assembly.
To the best of our knowledge, this is the first experimental study of {\it de novo} genome assembly problems both for real and synthetic data on quantum annealing devices and quantum-inspired techniques. 
\end{abstract}

\maketitle

Over the past few decades, an amount of DNA-related data has been increasing exponentially~\cite{Robinson2015}, and genomics is by now a data-driven science~\cite{Theis2019}.
More than 40 years ago, the first DNA genome ($\phi$X 174 bacteriophage) was sequenced~\cite{Sanger1977}. 
It took almost 13 years to sequence the whole human genome. 
Today, public and private facilities offer human genome sequencing that takes days or weeks~\cite{Park2016}. 
Current technologies sequence whole genomes in an unstructured set of reads with partial overlapping. 
However, the task of assembling DNA, {\it i.e.}, aligning and merging reads in order to reconstruct the original genome, 
which is a required step for most of the applications, still remain challenging~\cite{Liao2019}.

Existing approaches in sequencing read analysis are based on {\it de novo} assembling or mapping to an established reference.
{\it De novo} assembly is a method for constructing the original DNA sequence from the unstructured set of reads without any prior knowledge of the source DNA sequence length, layout, or composition~\cite{Chaisson2015}. 
{\it De novo} assembly is then essential for studying new species and structural genomic changes that cannot be detected by reading mapping. 
The complexity of {\it de novo} assembly depends on the genome size, abundance, length of repetitive sequences, and possible polyploidy~\cite{Chaisson2015}. 
For example, {\it de novo} assembly of a tiny $\phi$X 174 genome (5386 base pairs) on a laptop takes 10 minutes, while for the human genome ($3.2 \times 10^6$ base pairs) it takes about 48 hours on a supercomputer~\cite{Wong2018}. 
This time scale is acceptable in research tasks, but it is a limitation for emergency applications (including the clinical use).
Read mapping on a backbone of the reference genome is computationally more simple and allows detection of single- and oligonucleotide mutations, which are the major causes of human diseases~\cite{Lee2018}. 
However, the detection of genome rearrangements is a complicated task~\cite{Yao2019}. 
Read mapping algorithms used for the analysis of clinically important samples use local {\it de novo} assembly to correct mapping errors and reference mismatches~\cite{Li2019}. 
{\it De novo} assembly is currently used in transcriptome and cancer analysis, as gene fusions and genome rearrangements are common causes of malignant tumours~\cite{Miller2010}.
Decreasing the costs of sequencing makes whole-genome sequencing an irreplaceable part of personalized medicine and cancer treatment. 
The utility of sequencing technologies requires improved workflows with {\it de novo} assemblers to uncover significant genomic rearrangements in cancer and normal tissues.

Early generations of assembly tools are based on the overlap layout consensus (OLC) algorithm~\cite{Myers2005}.
Overlap discovery involves all-against-all, pair-wise read comparison, where one sets up the minimal number of shared nucleotides between two reads and an allowed number of mismatches. 
In the OLC graph, each read is represented by a vertex and an edge between two vertices indicates the overlap between corresponding reads. 
Then, finding the Hamiltonian path, i.e., the path that goes through all vertices and visits each vertex only once, allows reconstructing the original genome. 
This approach is widely used in assemblers (e.g., see Ref.~\cite{Myers2000}) and becomes suitable for the single-molecule sequencing technologies~\cite{Li2016, Koren2017}.

\begin{figure*}[htbp]
\includegraphics[width=1\linewidth]{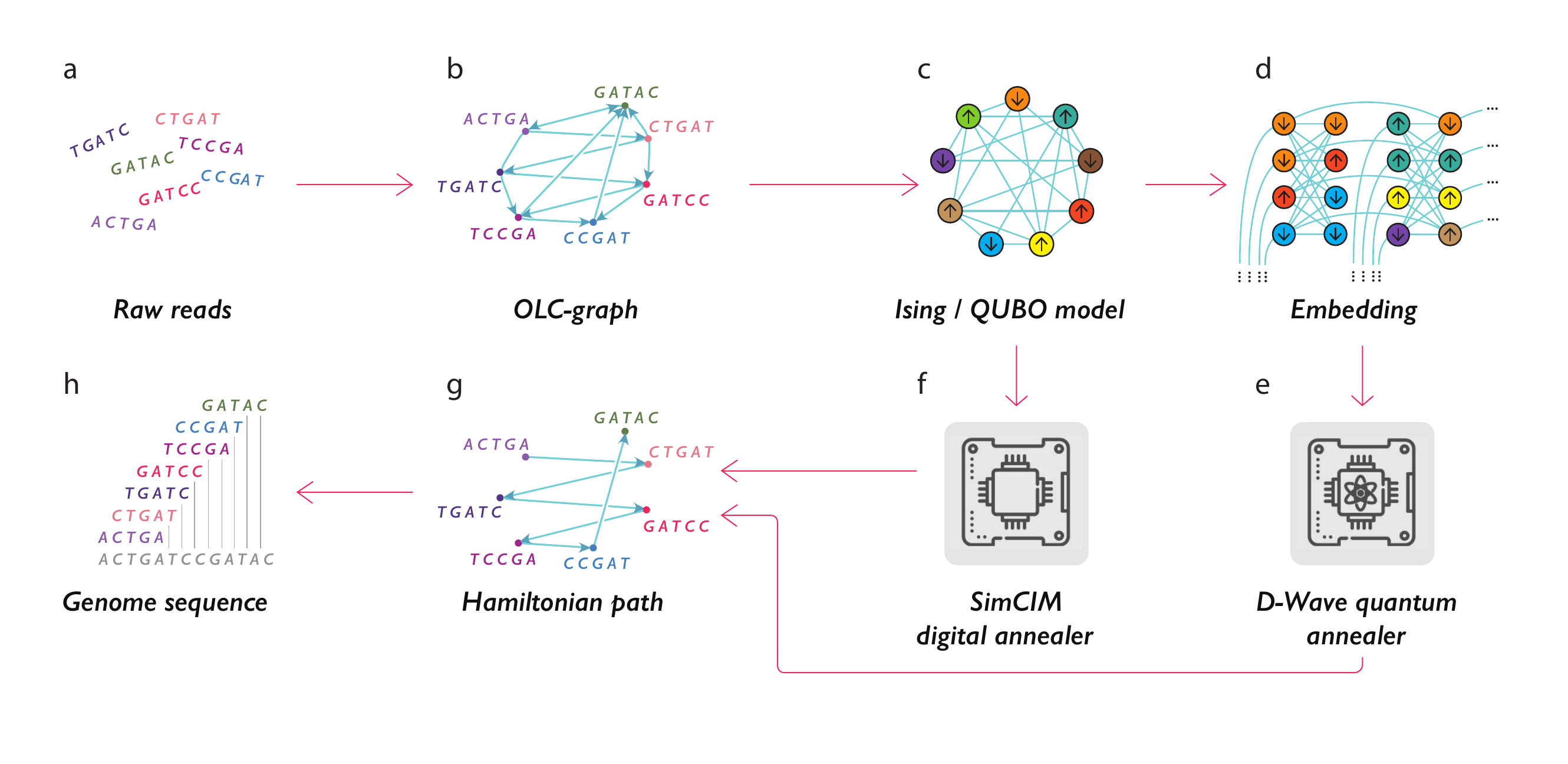}
\vskip -10mm
\caption{Solving the {\it de novo} genome assembly problem using quantum annealers and quantum-inspired (digital) annealing algorithms: 
a) raw reads; 
b) raw reads are transformed to the overlap-layout-consensus (OLC) graph; 
c) finding the Hamiltonian path for the OLC graph is reduced to the QUBO problem;
d) the QUBO problem should be embedded to the architecture of the quantum annealer (D-Wave): for this purpose each logical variable of the the QUBO problem is assigned with several qubits of the quantum annealer;
e) and f) the Ising problem in QUBO form can be solved using quantum annealers (D-Wave) and quantum-inspired algorithms (SimCIM), correspondingly; 
g) the output is the Hamiltonian path; 
h) the genome sequence is obtained as the solution.}  
\label{fig:scheme}
\end{figure*}

It is well known that finding a Hamiltonian cycle belongs to the class of NP-complete computational problems, for which finding an efficient solution is very hard. 
That is why the other graph representation is applied to the analysis of the sequencing data. 
This approach is related to the concept of De Bruijn graphs (DBG)~\cite{Tesler2011}. 
The idea is to construct a graph based on the fragmentation of reads down to smaller sequences called $k$-mers, where $k$ is the length of subsequence. 
These $k$-mers are aligned using $(k-1)$ sequence overlaps. 
Each node in the resulting $k$-mer graph represents a certain $k$-mer, while edges correspond to the overlaps between the $k$-mers. 
Thus, each read is represented by the sequence of the connected vertices (so-called overlapped $k$-mers). 
To obtain the original genome sequence one needs to find the Eulerian path, i.e., the path that visits each edge exactly once, but is allowed to revisit any vertex. 
The comparison of the OLC and DBG approaches are discussed in more detail in Ref.~\cite{Li2012}. 
As was mentioned above, the time of {\it de novo} assembling may be crucial for many applications. 
Possible improvements and speed-ups of the DBG approach, employing the Eulerian path, have been considered, 
whereas methods based on finding the Hamiltonian path in the OLC approach are less studied.

Quantum computers are a new generation of devices that use quantum phenomena, such as superposition and entanglement, for solving computational tasks. 
It is believed that quantum computers~\cite{Ladd2010} have a great potential to outperform existing technologies in various problems~\cite{Montanaro2017}, 
such as simulating complex systems~\cite{Lloyd1996}, machine learning~\cite{Biamonte2017}, and optimization~\cite{Farhi2016}. 
Ongoing research are related to the question how quantum computers could be used for computational biology and bioinformatics~\cite{Harrow2019}.
One of the algorithms that can be realized using quantum computers is Grover's search~\cite{Grover1996}, which can be used as a subroutine for sub-sequence alignment with a quadratic speed-up~\cite{Sarkar2019}.
There is increased activity at the interface of machine learning and quantum computing in the computational biology domain~\cite{Harrow2019,Prousalis2019,Fedorov2021}.
Being of extreme interest from the viewpoint of obtaining polynomial and exponential computational speedups, 
the suggested quantum algorithms~\cite{Sarkar2019,Grover1996,Harrow2019,Prousalis2019,Fedorov2021,Lindvall2019} require both a significant number of qubits and quite low error rates, 
which are beyond the capabilities of existing noisy intermediate-scale quantum (NISQ) devices.

Although many different implementations and models of quantum computing are in development, one of the approaches, which is based on quantum annealing, deserves a special attention.
This is due to the fact that the hardware for its implementation is available (it is produced by D-Wave System)~\cite{Boixo2013,Boixo2014,Ronnow2014}.
However, the ability of realistic quantum annealing devices to demonstrate computation speedups is still a subject of ongoing research~\cite{Vazirani2014,Ronnow2014,Katzgraber2014,Venturelli2015,Hen2015,Amin2015}. 
An interesting outcome of these debates is the appearance of a new generation of quantum-inspired (digital annealing) algorithms, 
which are essentially classical but appear as a result of analysing the role of quantum phenomena in solving computational tasks~\cite{Tiunov2019,Berloff2018,Lloyd2019}.
Results on the comparison between available quantum annealers and quantum-inspired algorithms on realistic optimization problems have been obtained~\cite{Lloyd2019}.  
We note that quantum annealing has been applied to various real-world tasks, including computational biology~\cite{Li2018}, exploration of the conformational landscape of peptides and proteins~\cite{Aspuru-Guzik2012}, 
and genome sequence alignment~\cite{Lindvall2019}.

Here we investigate the {\it de novo} genome assembly problem within the framework of quantum annealing. 
On the one hand, the OLC approach is less sensitive to sequencing errors and repeats than the DBG approach which leads to the higher quality of the assembly. 
On the other hand, the main disadvantage of the OLC-based algorithms is their computational inefficiency~\cite{Li2012}. 
Particularly, as we need to solve the NP-hard problem of finding a Hamiltonian path in the graph. 
Quantum computations have already shown their potential suitability for solving such problems. 
The above mentioned arguments motivate us to focus on the OLC formulation of the {\it de novo} genome assembly problem.
The main step in our study is to map the genome assembly problem in the framework of OLC graphs to a quadratic unconstrained binary optimization (QUBO) problem, 
which can be then efficiently embedded in the quantum annealing architecture (see Fig.~\ref{fig:scheme}). 
We also show that the genome assembly problem can be solved with the use of quantum-inspired optimization algorithms.
We note that our idea is to use quantum optimization for {\it de novo} sequencing, while other problems related to the analysis of genetic data are beyond the scope of the present work.
 
\section*{QUBO reformulation of the genome assembly problem}

The growing interest in the use of quantum computing devices (and in particular to quantum annealers) is related to their potential in solving combinatorial optimization problems.
It is widely discussed that the potential of quantum annealing is rooted in the quantum effects that allow us to efficiently explore the cost-function landscape in ways unavailable to classical methods.
Therefore, the important stage is to map the problem of interest to a Hamiltonian, which maps the binary representation of a graph path into a corresponding energy value.
The existing physical implementation of quantum annealing is the D-Wave processor, which can be described as Ising spin Hamiltonian.
The Ising Hamiltonian can be transformed into a QUBO problem. 
Thus, we have to find the mapping to a problem that we would like to solve on the D-Wave quantum processor to the QUBO form. 
However, establishing correspondence between a problem of interest and the QUBO form may require additional overhead. 
In particular, in our case the transformation of the OLC graph to a QUBO problem requires the use of additional variables (see below).

In addition, the D-Wave quantum processor has its native structure (the chimera structure).
That is why after the formulation of the problem of interest in the QUBO form an additional stage of embedding problem in the native structure of the quantum device is required.  
So additional overhead in the number of physical qubits, which is related to the representation of logical variables by physical qubits of the processor (that takes into account the native structure), is required (see below). 

\subsection*{Formulation of the Hamiltonian path problem}

Along the lines of Ref.~\cite{Lucas2014}, we reformulate the task of finding the Hamiltonian path in the OLC graph as a QUBO problem.

Let a directed OLC graph be given in the form $G = (V, E)$, where $V=\{1,2,\ldots,N\}$ is a set of vertices, and $E$ is set of edges consisting of pairs $(u,v)$ with $u,v\in V$.
The solution of the Hamiltonian path problem is represented in the form of $N\times N$ permutation matrix ${\cal X}=(x_{v,i})$, whose unit elements $x_{v,i}$ represent the path going through the $v$th node at the $i$th step. 
Then, we assign {\it each} element $x_{v,i}$ of the matrix ${\cal X}$ a separate logical variable (spin) within an optimization problem.
Note, that this representation results in polynomial overhead in the number of logical variables of the QUBO problem: The solution for $N$-vertex graph requires $N^{2}$ logical variables.
The resulting Hamiltonian of the corresponding QUBO problem takes the following form:
\begin{equation}\label{eq:QUBO}
	\begin{split}
		\mathcal{H}&=A \sum_{v=1}^{N}\left(1- \sum_{j=1}^{N}x_{v,j}\right)^2 \\
		& + A \sum_{j=1}^{N}\left(1- \sum_{v=1}^{N}x_{v,j} \right)^2 \\
		& + A \sum_{(u,v)\notin E}\sum_{j=1}^{N-1}(x_{u,j}x_{v,j+1}),
	\end{split}
\end{equation}
where $A>0$ is a penalty coefficient.
The first two terms in Eq.~\eqref{eq:QUBO} ensure the fact that each vertex appears only once in the path, and there is a single vertex at each step of the path.
The third term provides a penalty for connections within the path that are beyond the allowed ones.
With this QUBO formulation, we are able to run the genome assembly task using quantum annealers and quantum-inspired algorithms.
We note that the applicability of the method requires the existence of the Hamiltonian path in the corresponding graph, which is not universally the case for arbitrary genetic data given by an OLC-graph.  

\subsection*{Formulation of the Hamiltonian path problem for acyclic graphs}

In general, Hamiltonian path mapping is suitable both for cyclic and acyclic directed graphs. 
However, it is often the case that the OLC graph contains no cycles. 
It is then possible to further simplify transformation and reduce the qubit overhead. 
Here, we demonstrate more compact mapping that requires only $M$ logical variables, where $M=|E| < N^{2}$ is the number of edges.
For the acyclic OLC graph $G = (V, E)$, let us define a set of binary variables $\{x_{u,v}\}_{(u,v)\in E}$ that indicate whether an edge $(u,v)$ is included in the Hamiltonian path.
Then the corresponding Hamiltonian should include the following two components:
\begin{equation}\label{formula:hamiltonian_long}
\begin{split}
	\mathcal{H}= A
	\sum\limits_{\substack{u\in V}}&\left(1-\sum_{\substack{(u,v)\in E}}x_{u,v} \right)^2 \\
	&+A\sum\limits_{\substack{v\in V}}\left(1-\sum_{\substack{(u,v)\in E}}x_{u,v} \right)^2.
\end{split}
\end{equation}

The first and the second terms in Eq.~\eqref{formula:hamiltonian_long} assure that each vertex is incident (if possible) with a single incoming and outgoing path edges correspondingly.
Although this realization is helpful and can be used for solving genome assembly problems on quantum annealers without polynomial qubit overhead 
(the encoding requires $M$ variables, where $M$ is the number of edges in the corresponding OLC-graph), 
the asymptotic computational speed-up versus classical algorithm is not exponential (since there exist efficient classical algorithms for solving this problem).

\subsection*{Embedding to the processor native structure} 

As it is mentioned before, the logical variables in Hamiltonians~\eqref{eq:QUBO} and \eqref{formula:hamiltonian_long} are not necessarily equivalent to physical qubits of a quantum processor. 
This is due to the fact that quantum processors have its native structure, i.e. topology of physical qubits and couplers between them.
For example, each D-Wave 2000Q QPU is fabricated with 2048 qubits and 6016 couplers in a C16 Chimera topology~\cite{Boixo2013,Boixo2014,Ronnow2014}. 
Within a C16 Chimera graph, physical qubits are logically mapped into a $16\times16$ matrix of unit cells of $8$ qubits, where each qubit is connected with at most 6 other qubits.
In order to realize Hamiltonians~\eqref{eq:QUBO} and~\eqref{formula:hamiltonian_long}, whose connections structure may differ from the one of the processor, we employ an additional embedding stage, described in~\cite{Zbinden2020}.
It allows obtaining a desired effective Hamiltonian by assigning several physical qubits of the processor to a single logical variable of the original Hamiltonian (see Fig.~\ref{fig:scheme}c-d).
The embedding to the native structure introduces considerable overhead in qubit number relative to the fully-connected model, yet allows solving problems with existing quantum annealers.

\section*{Results}

Here we apply our method for the experimental realization of {\it de novo} genome assembly using quantum and quantum-inspired annealers.  
As a figure of merit we use the time-to-solution (TTS), which is the total time required by the solver to find the optimal solution (ground state) at least once with a probability of 0.99.
We first define $R_{99}$ as the number of runs required by the solver to find the ground state at least once with a probability of 0.99.
Using binomial distribution one can calculate $R_{99}$ as follows:
\begin{equation}
	R_{99} = \frac{\log(1-0.99)}{\log(1-\theta)},
\end{equation}
where $\theta$ is an estimated  success probability of each run.

Then we define TTS it in the following way:
\begin{equation}
	{\rm TTS} = {t_{a}}R_{99},
\label{formula:tts}
\end{equation}
where $t_{a}$ for D-Wave is 20$\mu{s}$ (default value). 

Quantum-inspired optimization algorithms can be also used for solving QUBO problems.
In our experiments, we employ SimCIM quantum-inspired optimization algorithm~\cite{Tiunov2019}, 
which is based on the differential approach to simulating specific quantum processors called Coherent Ising Machine (CIM; see Methods).
SimCIM runs on conventional hardware and is easily parallelizable on graphical processing units (GPU). 
This is the time for simulating a single annealing run using our implementation of SimCIM, measured on Intel core i7-6700 Quad-Core, 64GB DDR4, GeForce GTX 1080.

\subsection*{$\phi$X~174 bacteriophage genome}

We start with the paradigmatic example of the $\phi$X~174 bacteriophage genome~\cite{Sanger1977}.
In order to realize {\it de~novo} genome assembly, we construct the adjacency matrix for OLC graphs and use pre-processing for packing this graph into D-Wave processor.
We then transform each adjacency graph into the QUBO matrix according to Eq.~(\ref{formula:hamiltonian_long}).
For the case of using the D-Wave system we embed the QUBO problem in the native structure of the annealing device (which naturally adds overhead in the number of qubits; see Methods).
In order to embed $\phi$X 174 graph into the D-Wave processor, we have preliminary conducted manual graph partitioning using classical algorithms implemented in the METIS tool~\cite{METIS}.
Then the problem can be solved with the use of quantum annealing hardware by D-Wave and quantum-inspired optimization algorithm.

For each instance, a total of 10$^3$ anneals (runs) were collected from the processor, with each run having an annealing time of 20 $\mu$s. 
The total number of instances is 1000 (the process of their generation is described in Methods).
The results are presented in Table~\ref{tab:bacteriophage}.
Up to our best knowledge, this is the first realistic-size {\it de novo} genome assembly employing the use of quantum computing devices and quantum-inspired algorithms.
The presented time is required for finding the optimal solution since only one solution has a right interpretation.
We note that the time required for the data pre-processing is not included in Table~\ref{tab:bacteriophage}.
Details of graph size for each part after manual graph partitioning is presented in Table~\ref{tab:graph} in Methods.

Statistics presented for 1000 simulated  Phi-X 174 bacteriophage OLC graphs. 
We use D-Wave hybrid computing mode, which employs further graph decompositions with parallel computing on both classical and quantum backends. 
D-Wave gives access to the following timing information in the information system: $T_{\rm run}$ (run time), $T_{\rm charge}$ (charge time), and $T_{\rm QPU}$ (QPU access time).
We assume CPU time $=T_{\rm run}-T_{\rm QPU}$.  
We summarize QPU access time and CPU time for obtained OLC graphs. 
For the case of SimCIM, we use TTS.

\begin{table}[]
\begin{tabular}{|l|l|l|l|l|l|}
\hline
                                                                                   &     & Mean, $\mu$s & Min, $\mu$s & Max, $\mu$s & 90\% \\ \hline
\multirow{2}{*}{\begin{tabular}[c]{@{}l@{}}Quantum\\ annealer\end{tabular}}        & CPU & 8483         & 8314        & 8619        & 8579 \\ \cline{2-6} 
                                                                                   & QPU & 535          & 369         & 672         & 600  \\ \hline
\begin{tabular}[c]{@{}l@{}}Quantum-\\ inspired\\ annealer \\ (SimCIM)\end{tabular} &     & 262          & 9.9         & 7212        & 1061 \\ \hline
\end{tabular}
\caption{Genome assembly time for $\phi$X 174 bacteriophage for 1000 instances. 
For the data based on experiments with quantum annealers we highlight required classical processor unit (CPU) time and quantum processor unit (QPU) time.}
\label{tab:bacteriophage}
\end{table}

\subsection*{Benchmarking quantum-assisted {\it de novo} genome assembly using the synthetic dataset}

In order to perform a complete analysis of the suggested approach, we realize the quantum-assisted {\it de novo} genome assembly for the synthetic dataset.
We generate a synthetic dataset, which consists of 60 random reads of length from 5 to 10 (for details, see Methods), 10 problems are generated for every sequence length. 
We then split each read into $k$-mers of length 3 and compute adjacency matrix for the corresponding OLC graph using Eq.~(\ref{formula:hamiltonian_long}).
Finally, we transform each adjacency graph into the QUBO matrix according to our algorithm and minimize it using quantum annealing hardware by D-Wave and quantum-inspired optimization algorithms.

\begin{figure}
	\includegraphics[width=1\linewidth]{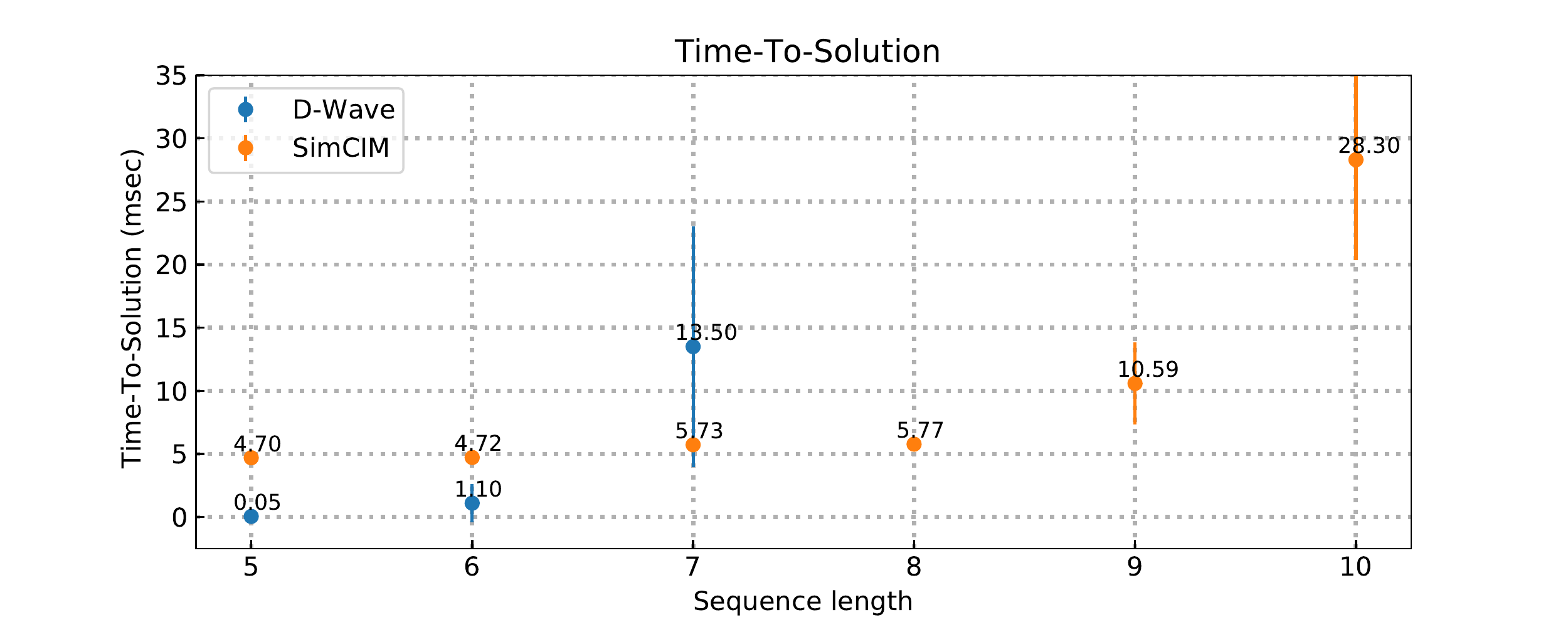}
	\vskip -4mm
	\caption{Comparison of the performance of quantum and quantum-inspired methods for {\it de novo} genome assembly based on synthetic data (10 problems were generated for every sequence length): 
	we compare TTS  for quantum device D-Wave and quantum-inspired optimization algorithm SimCIM.}
	\label{fig:comparison}
\end{figure}

Our goal is to check the applicability of existing quantum annealers to the task of genome assembly, 
evaluate the upper bound on the input problem size (particularly, the length of the original genome), compare the performance of the D-Wave quantum annealer with a software annealing simulator SimCIM. 
The choice of tools is motivated by their maturity in terms of quantum dimensionality and compatibility with the original formulation in terms of the optimization problem.
The similar routine is realized with the use of quantum-inspired annealing.
We test the suggested approach with the simulated data first with the D-Wave quantum annealer (see Fig.~\ref{fig:comparison}) and compare our results with quantum-inspired optimization algorithm SimCIM.

We note that the D-Wave annealer shows an advantage in genome assembly for short-length sequences, 
while it cannot be applied for sequences of length 6 and more due to the fact that the decoherence time becomes comparable with the annealing time. 
What we observe is that the performance of the annealing system is dependent on the properties of input data. 
While the exhaustive investigation of the nature of D-Wave performance with respect to input data goes beyond the scope of our research, we consider the connectivity of the input graph as one of the critical factors. 
The D-Wave 2000Q processor is based on Chimera Topology with native physical connectivity of 6 basic qubit cells. 

During the experiments on the synthetic dataset, we were able to embed the problems with a maximum sequence of up to 10 nucleotides. 
This length corresponds to the fully connected graph with 8 nodes (K8,8) and this is the maximum possible graph size, which can be embedded to the chimera lattice using clique embedding tool from DWave Ocean SDK. 
While Ocean SDK allows using other types of embedding (e.g., minor miner) we observed that clique embedding demonstrates more stable results due to the deterministic nature of embedding in comparison to the minor-miner tool, 
which is intrinsically based on randomized heuristics contributing to larger deviations across experiments. 
However, no viable solution that could reconstruct the original sequence was found for sequences longer than 6.

\section*{Discussions}

In our work, we have demonstrated the possibility of solving the simplified bioinformatics problem of reconstructing genome sequences using quantum annealing hardware and quantum-inspired algorithms.
We have implemented the experimental quantum-assisted assembly of $\phi$X 174 bacteriophage genome.
On the basis of synthetic data, we have shown that the existing quantum hardware allows reconstructing short sequences of up to 7 nucleotides. 
In order to use quantum optimization for realistic tasks, the ratio of the decoherence time to the annealing time should be considerably improved. 
We note that while the decoherence time is not a fundamental limitation of the technology, the realization of quantum annealers with sufficient decoherence time remains a challenge.
While D-Wave machines use superconducting quantum circuits, 
setups based on ultracold Rydberg atom arrays~\cite{Lukin2017,Lukin2018,Browaeys2020} and trapped ions~\cite{Monroe2017,Blatt2018} can be also used for the efficient implementation of quantum annealing and other quantum optimization algorithms. 
Specifically, the system of Rydberg atom arrays has been studied in the context of solving the maximum independent set problem~\cite{Lukin2018,Browaeys2020}, which is NP-hard.
For longer sequences, as we have demonstrated, it is possible to use quantum-inspired algorithms that are capable of solving more complex problems using classical hardware. 

We note that our work is a proof-of-principle demonstration of the possibility to use existing quantum devices for solving the genome assembly problem. 
The problem scale considered in this paper is still far from real sequences ($\sim$130 kilo-base pairs for primitive bacterias) and is lacking numerous complications, such as errors in sequence reads and handling repeating sequences. 
However, the proposed method demonstrates that newly evolving computing techniques based on quantum computers and quantum-inspired algorithms are quickly developing and can be soon applied in new areas of science.

Limitations of existing quantum hardware do not allow universally outperform existing solutions for {\it de novo} genome assembling. 
At the same time, one of the most interesting practical questions is when one can expect computational advantages from the use of quantum computing in genome assembling tasks. 

We note that in real-life conditions a number of additional challenges arise.
Examples include errors (random insertions and deletions, repeats, etc.), genome contaminants (pieces of the genome not related to the subject of interest), polymer chain reaction artifacts, and others require additional post-processing steps. 
These problems are beyond the scope of our proof-of-principle demonstration and they should be considered in the future. 
Another complication comes from the fact that temperature and other noise effects play a significant role in the case of the use of realistic quantum devices.
Thermal excitation and relaxation processes affect performance. 
Our further directions include optimization of the QUBO model for more compact spin representation and integration of error model into our algorithm. 
Solving these two issues can enable reconstruction of real sequences using the quantum approach.

%Our source code for a proof-of-principle realization of the quantum-assisted genome assembly is freely available under the GNU general public license.
%The realization also contains the synthetic set that can be used for reproducing the obtained results. 
%For further details of the implementation, we refer the reader to the description of D-Wave's Software.

\section*{Methods}

\subsection*{Quantum annealing protocol}

The beginning Hamiltonian of the D-Wave processor is a transverse magnetic field of the following form:
\begin{equation}
	\mathcal{H}_0=\sum_{i\in{V}}h_i\sigma_i^x,
\end{equation}
where $\sigma_i^x$ is the Pauli $x$-matrix, which acts on $i$th qubit.
The problem Hamiltonian can be encoded to the following Ising Hamiltonian:
\begin{equation}
	\mathcal{H}_{\rm P}=\sum_{i\in{V}}h_i\sigma_i^z+\sum_{(i,j)\in{E}}J_{ij}\sigma_i^z\sigma_j^z,
\end{equation}
where $h_i$ describe local fields, $J_{ij}$ stands for couplings, $\sigma_i^z$ are the Pauli $z$-matrices, and $E$ is the set of edges. 
One can see that $\mathcal{H}_{\rm P}$ is of diagonal form, so $\sigma_i^z$ can be treated as spin values $\{\sigma_i^z=\pm1\}$.
For a given spin configuration ${\sigma_i^z}$ the total energy of the system is given by $\mathcal{H}_{\rm P}$, so by measuring the energy one can find a solution to the problem of interest. 

Quantum annealing can be applied to any optimization problem that can be expressed in the QUBO form.
The idea is then to reduce the problem of interest to the QUBO form.

\subsection*{QUBO transformation}

The Ising Hamiltonian can be directly transformed to a quadratic unconstrained binary optimization (QUBO) problem. 
The following transformation can be applied for this purpose:
\begin{equation}
	w_i=\frac{\sigma_i^z+1}{2} \in \{0,1\},
\end{equation}
where $\{\sigma_i^z=\pm1\}$.
For solving the problem on the D-Wave quantum processor, all $h_i$ and $J_{ij}$ values are scaled to lie between $-1$ and $1$.
As a result, the processor outputs a set of spin values $\{\sigma_i^z=\pm1\}$ that attempts to minimize the energy, and the lower energy indicates better solution of the optimization problem. 
We note that Ref.~\cite{Lucas2014} provides a method for QUBO/Ising formulations of many NP problems.

\subsection*{Quantum-inspired annealing using SimCIM}

SimCIM is an example of a quantum-inspired annealing algorithm, which works in an iterative manner.
It can be used for sampling low-energy spin configurations in the classical Ising model. 
The algorithm treats each spin value $s_i$ as a continuous variable, which lie in the range $[-1, 1]$.
Each iteration of the SimCIM algorithm starts with calculating the mean field 
\begin{equation}
	\Phi_i = \sum_{j \neq i}J_{ij}s_j + b_i, 
\end{equation}
which act on each spin by all other spins ($b_i$ is an element of the bias vector). 
Then the gradients for the spin values are calculated according to $\Delta s_i = p_t s_i + \zeta \Phi_i + N(0,\sigma)$,
 where $p_t, \zeta$ are the annealing control parameters and  $N(0,\sigma)$ is the noise of the Gaussian form. 
 Then the spin values are updated according to $s_i \leftarrow \phi(s_i + \Delta s_i)$, where $\phi(x)$ is the activation function
\begin{equation}\label{acti}
	\phi(x)=\begin{cases}
	x \textrm{ for } |x|\leq 1;\\
	x/|x| \textrm{ otherwise}
\end{cases}
\end{equation}
After multiple updates, the spins will tend to either $-1$ or $+1$ and the final discrete spin configuration is obtained by taking the sign of each $s_i$. 

\subsection*{Bacteriophage simulations}

\begin{figure}[t]
	\includegraphics[width=0.65\linewidth]{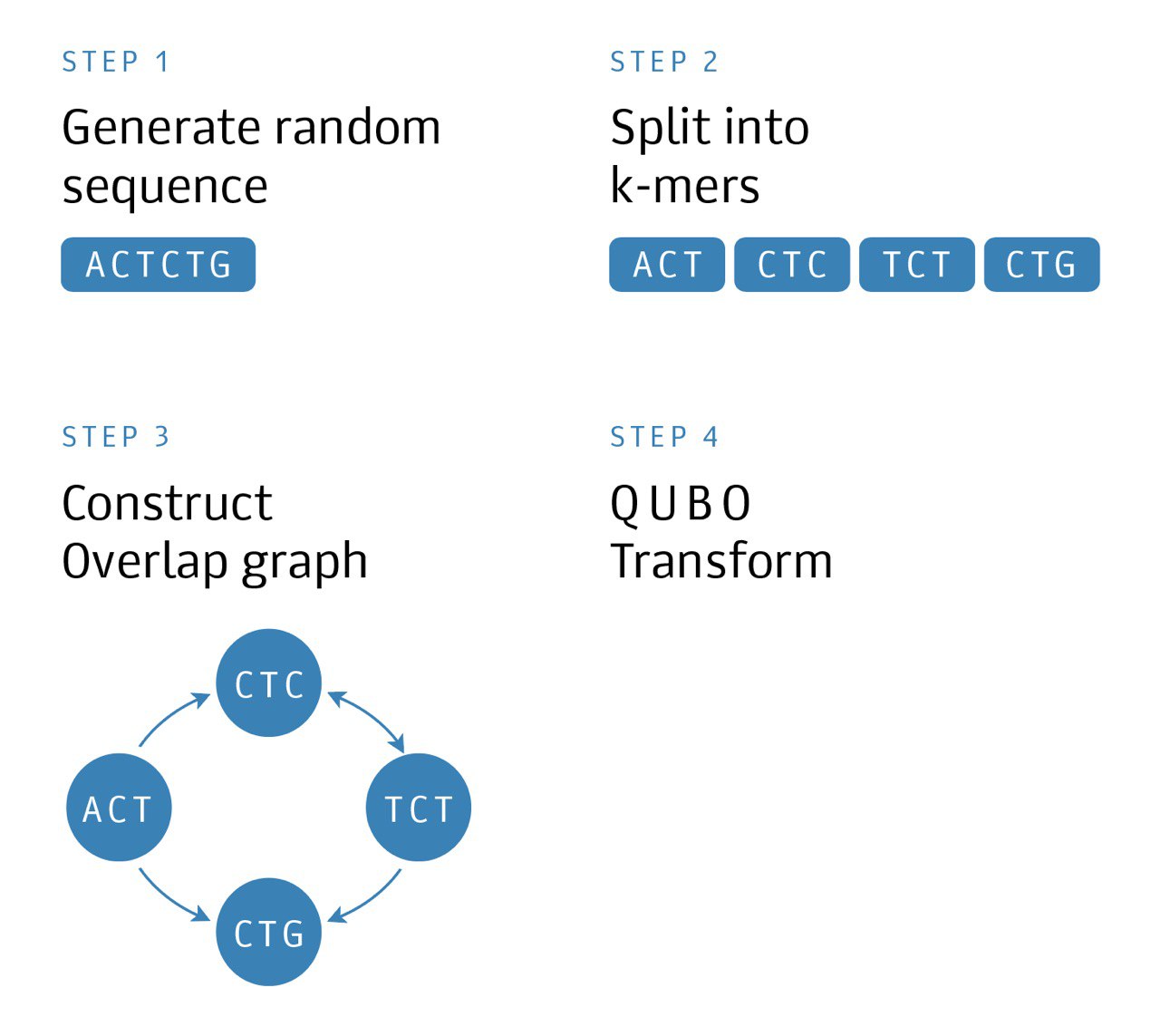}
	\vskip -4mm
	\caption{Experimental scheme for the synthetic dataset.}
	\label{fig:dataset}
\end{figure}

We use {\it Grinder} \cite{Angly2012} to simulate raw reads from $\phi$X 174 bacteriophage complete genome (NCBI Reference Sequence: NC$\underline{\quad}$001422.1). 
To simplify the task and make it feasible for quantum computing we generate 50 reads in each run of simulations. 
In our proof-of-concept research, we are focused on finding the Hamiltonian path in OLC graph using quantum and quantum-inspired annealing. 

We generate the raw reads with no sequencing errors and the length of each read is equal to 600 base pairs. 
We build the OLC graph using the pairwise alignment of the raw reads implemented in {\it minimap2} package~\cite{Li2018_2}. 
We run {\it minimap2} with the predefined set of parameters {\sf ava-ont} and $k=10$. 
We apply {\it miniasm}~\cite{Li2016} to the same data as the benchmark assembler, which uses OLC graphs.

\medskip
\begin{table}[]
\begin{tabular}{|c|c|c|c|}
\hline
Sequence length & Graph size & Qubo size & Physical qubits \\ \hline
5               & 3          & 9         & 36             \\ \hline
6               & 4          & 16        & 80             \\ \hline
7               & 5          & 25        & 200            \\ \hline
8               & 6          & 36        & 360            \\ \hline
9               & 7          & 49        & 686            \\ \hline
10              & 8          & 64        & 1088           \\ \hline
\end{tabular}
	\caption{Experimental scheme for the synthetic dataset.}\label{tab:graph}
\end{table}

For experiments with quantum annealing, we use public access to D-Wave 2000Q via Leap SDK. 
We evaluate the impact of tunable parameters (particularly, annealing time) on the final solution quality; however, no significant improvement was discovered against default values, so annealing time was set to 20 $\mu$s (default value). 
The number of annealing runs is set to $10^3$ (maximum possible value).
During our experiments we use mostly the standard configuration of the D-Wave processor, so we do not have any specific requirements on the weights/couplers in the model.

Synthetic dataset graphs, which consist of reads no longer than 7 nucleotides (25 graph nodes), 
are small enough to fit into quantum annealer, so we can use DW$\underline{\quad}$2000Q$\underline{\quad}$5 backend (pure quantum mode of operation; see the following section).
However, the size of the $\phi$X 174 bacteriophage graph (248 vertices) is too large.
In order to embed $\phi$X 174 graph into the D-Wave processor, we have preliminary conducted manual graph partitioning using classical algorithms implemented in the METIS tool~\cite{METIS}. 
It allows splitting the original graph into 3 sub-parts. 
They are carefully selected so that only a single edge remains between them. 
The longest path is then calculated separately for each part and concatenated into a single path of the original graph. 
Each part is still large enough to be computed using the purely quantum mode, so we use the D-Wave hybrid computing mode --- hybrid$\underline{\quad}$v1 backend. 
D-Wave hybrid computing mode employs further graph decompositions with parallel computing on both classical and quantum backends. 
Specifics of such decomposition are not publicly available and physical qubit count is also not shown to the end-user. 
Details of graph size for each part after manual graph partitioning is presented in Table~\ref{tab:graph}.
According to D-Wave Leap specification, hybrid$\underline{\quad}$v1 backend automatically combines the power of classical and quantum computation.

\subsection*{Simulations with synthetic dataset}

In order to evaluate the performance of the algorithm in a controlled setup, we generated several hundreds of random nucleotide sequences with variable length and performed corresponding transformations as shown in Fig.~\ref{fig:dataset}. 
Further, we eliminated graph duplicates or other trivial cases, where graph structure contained no auxiliary edges. 
Synthetic dataset graphs up to the length of 7 nucleotides (25 graph nodes) are small enough to fit into quantum annealer, so we can use DW$\underline{\quad}$2000Q$\underline{\quad}$5 backend (pure quantum mode of operation).
Finally, we selected 60 sequences that produce unique OLC graphs with comparable complexity.

{\it Note Added}. Recently, we became aware of the work reporting studies of the quantum acceleration using gate-based and annealing-based quantum computing~\cite{Sarkar2020}.

Corresponding author: A.K.F. (akf@rqc.ru).

\section*{Acknowledgements}

We are grateful to A.I. Lvovsky for fruitful discussions as well as A.S. Mastiukova and D.V. Kurlov for useful comments. 
We also thank the anonymos referee for careful reading our manuscript and insightful comments that helped to improve the paper.
We thank E.S. Tiunov for providing information about the SimCIM algorithm and A.E. Ulanov for the discussion of various quantum-inspired optimization algorithms. 
This work is supported by Russian Science Foundation (19-71-10092).
We also thank D-Wave Systems (the research is conducted within the program of global response to COVID-19). 

\section*{Competing interests}

Owing to the employments and consulting activities of A.S.B., S.R.U., E.O.K., and A.K.F., they have financial interests in the commercial applications of quantum computing.
A.S.R., I.V.P., and V.V.I. are employees of Genotek Ltd, they declare that they have no other competing interests.
A.N.K. declares no competing interests.

\end{document}